\documentclass
[aps,preprintnumbers,prb,twocolumn,superscriptaddress,showpacs]{revtex4}%
\usepackage{graphicx}
\usepackage{dcolumn}
\usepackage{bm}
\usepackage[colorlinks=true,linkcolor=blue,anchorcolor=blue,citecolor=blue,urlcolor=blue]%
{hyperref}
\usepackage{multirow}
\usepackage{amsmath}
\usepackage{amsfonts}
\usepackage{amssymb}%
\providecommand{\U}[1]{\protect \rule{.1in}{.1in}}

\begin{document}
\title{Quantum Hall Effects in a Non-Abelian Honeycomb Lattice}
\author{Ling Li}
\affiliation{School of Science, Hebei University of Science and Technology, Shijiazhuang
050018, China}
\author{Ningning Hao}
\affiliation{Department of Physics, The University of Hong Kong, Pokfulam Road, Hong Kong, China}
\author{Guocai Liu}
\email{guocailiu@semi.ac.cn}
\affiliation{School of Science, Hebei University of Science and Technology, Shijiazhuang
050018, China}
\author{Zhiming Bai}
\affiliation{School of Science, Hebei University of Science and Technology, Shijiazhuang
050018, China}
\author{Zai-Dong Li}
\affiliation{Department of Applied Physics, Hebei University of Technology, Tianjin 300401, China}
\author{Shu Chen}
\affiliation{Beijing National Laboratory for Condensed Matter Physics, Institute of
Physics, Chinese Academy of Science, Beijing 100190, China}
\author{W. M. Liu}
\affiliation{Beijing National Laboratory for Condensed Matter Physics, Institute of
Physics, Chinese Academy of Science, Beijing 100190, China}

\begin{abstract}
We study the tunable quantum Hall effects in a non-Abelian honeycomb optical
lattice which is a multi-Dirac-point system. We find that the quantum Hall
effects present different features with the change in relative strengths of
several perturbations. Namely, the quantum spin Hall effect can be induced by
gauge-field-dressed next-nearest-neighbor hopping, which together with a
Zeeman field can induce the quantum anomalous Hall effect characterized by
different Chern numbers. Furthermore, we find that the edge states of the
multi-Dirac-point system represent very different features for different
boundary geometries, in contrast with the generic two-Dirac-point system. Our
study extends the borders of the field of quantum Hall effects in a honeycomb
optical lattice with multivalley degrees of freedom.

\end{abstract}

\maketitle

\section{Introduction}

The honeycomb lattice, which is the brick that builds graphene, plays a
significant role in promoting new physics. The most remarkable feature of the
energy band of the honeycomb lattice system is that the low-energy excitations
display a linear dispersion relation \cite{Wallace} and are, thus, described
by massless Dirac fermions \cite{Semenoff}. Furthermore, massless Dirac
fermions can be tuned into massive Dirac fermions by tunable perturbations.
More importantly, the transition from the gapless to the gapped phase can
result in new physics, such as the quantum anomalous Hall (QAH) effect in the
well-known Haldane graphene model \cite{Hald} and quantum spin Hall (QSH)
effect in the Kane-Mele graphene model \cite{Kane}. The former model pioneers
a new route to realize non-zero integer Hall conductance without Landau levels
and the latter model establishes the foundation to breed topological
insulators \cite{Hasan,Qi}. Comparison between the two models indicates the
power of the internal spin degrees of freedom to produce new physics in
honeycomb lattice systems.

On the other hand, the development of laser and ultracold atom techniques
gives rise to various optical lattices in which ultracold Fermi atoms can be
trapped to simulate the phenomena in condensed-matter systems
\cite{Anglin,Jaksch,Duan,Uehl,Liu,Liu1,Hao1}, such as the topological Mott
insulator \cite{Mathias2015SP} (MI) and superfluid (SF)--MI transition
\cite{Strohmaier,Schneider}. More significantly, in a laser field with a
specific configuration, the trapped ultracold atoms can feel effective Abelian
or non-Abelian gauge fields \cite{Osterloh,Jotzu}. The couplings between the
trapped ultracold atoms and the artificial gauge fields are equivalent to
various interactions in cold-atom systems, such as spin-orbital couplings
\cite{Wang2012PRL,Cheuk2012PRL}.

In this paper, we study a non-Abelian honeycomb optical lattice which is a
two-dimensional multi-Dirac-point system. Compared with the usual honeycomb
lattice, in which only two independent Dirac points emerge, the non-Abelian
honeycomb optical lattice here has eight Dirac points. Namely, the dimension
of the internal valley degrees of freedom is extended from 2 to 8, and the
extended valley degrees of freedom are induced by the couplings between
trapped atoms and the non-Abelian gauge fields. Likewise, we introduce three
kinds of perturbations to open a gap of Dirac points independently.
Explicitly, two perturbations including staggered sublattice potentials and
gauge-field-dressed next-nearest-neighbor (NNN) hopping are time-reversal
invariant. The third is Zeeman splitting, which breaks the time-reversal
symmetry of the system. We find that staggered sublattice potentials give a
trivial gapped phase with no gapless edge states in the ribbon geometry, and
the gauge-field-dressed NNN hopping results in the QSH effect, with the spin
direction lying in the ribbon plane. Interestingly, we find that the edge
states connecting different internal valleys present different features even
though the overall Chern number is 0 in the presence of both staggered
sublattice potentials and Zeeman splitting. Furthermore, it is shown that
cooperations among the three kinds of perturbations can induce QAH effects
characterized by different Chern numbers. Likewise, edge states show internal
structures for different types of edge boundaries. The results indicate that
many-Dirac-point systems represent more abundant physics in comparison with
the simple two-Dirac-points systems, and the robustness of the bulk-boundary
correspondence is global, not local.

The paper is organized as follows. In Sec. II, we introduce the non-Abelian
honeycomb lattice model with three kinds of perturbations. In Sec. III, we
study the topological phase transition driven by three kinds of perturbations
by investigating the change in the valley Chern numbers defined at Dirac
points. Furthermore, we study the relations between the change in Chern
numbers and the emergence of edge states under different geometries. In Sec.
IV, we discuss some important relevant issues and give a brief summary.

\section{Model}

We start with a non-Abelian honeycomb lattice model proposed in Ref.
\cite{Berm} and introduce three kinds of perturbations into this model to open
a gap at each Dirac point, respectively. The tight-binding Hamiltonian takes
the form,%
\begin{equation}
H=H_{0}+H_{1}, \label{E0}%
\end{equation}
with%
\begin{align}
H_{0}  &  =-t\sum_{\langle i,j\rangle}b_{j}^{\dag}U_{ij}a_{i}+H.c.,
\label{E1}\\
H_{1}  &  =-t_{1}\sum_{\langle \langle i,j\rangle \rangle}\left(  a_{j}^{\dag
}U_{ij}^{\prime}a_{i}+b_{j}^{\dag}U_{ij}^{\prime}b_{i}\right) \nonumber \\
&  +\delta \sum_{i}\left(  a_{i}^{\dag}a_{i}-b_{i}^{\dag}b_{i}\right)
+h\sum_{i}\left(  a_{i}^{\dag}\sigma_{z}a_{i}+b_{i}^{\dag}\sigma_{z}%
b_{i}\right)  . \label{E2}%
\end{align}
Here $t$ ($t_{1}$) is the hopping amplitude between the (next) nearest
neighbor link $\langle i,j\rangle$ ($\langle \langle i,j\rangle \rangle$),
$a_{i}^{\dag}$=$\left(  a_{i\uparrow}^{\dag},a_{i\downarrow}^{\dag}\right)  $
with $a_{i\alpha}^{\dag}$ ($a_{i\alpha}$) denoting the creation (annihilation)
operator of a fermionic atom with spin $\alpha$ (up or down) at A sublattice
$i$ (a similar definition applies for sublattice B). The unitary operator
$U_{ij}$ ($U_{ij}^{\prime}$) is associated with the link connecting the (next)
nearest-neighbor lattice points $\mathbf{r}_{i}\rightarrow \mathbf{r}_{j}$
\cite{Berm}. The unitary operators coupling fermionic atoms to non-Abelian
fields along each hopping path from sublattice to sublattice B are taken as
$U_{1}=e^{i\alpha_{1}\sigma_{x}}$, $U_{2}=1$, $U_{3}=e^{i\alpha_{3}\sigma_{y}%
}$. The three nearest neighbor hopping pathes are $\mathbf{d}_{1}=a\left(
\frac{\sqrt{3}}{2},-\frac{1}{2}\right)  $, $\mathbf{d}_{2}=a\left(
0,1\right)  $, $\mathbf{d}_{3}=a\left(  -\frac{\sqrt{3}}{2},-\frac{1}%
{2}\right)  $. Similarly, the unitary operators along each hopping path
between the NNN lattice points are chosen as $U_{1}^{\prime}=U_{4}^{\prime
\dag}=e^{i\gamma_{1}\sigma_{z}}$, $U_{2}^{\prime}=U_{5}^{\prime \dag
}=e^{i\gamma_{2}\sigma_{z}}$, $U_{3}^{\prime}=U_{6}^{\prime \dag}%
=e^{i\gamma_{3}\sigma_{z}}$, with NNN hopping paths $\mathbf{b}_{1}%
=-\mathbf{b}_{4}=a\left(  \sqrt{3},0\right)  $, $\mathbf{b}_{2}=-\mathbf{b}%
_{5}=a\left(  \frac{\sqrt{3}}{2},\frac{3}{2}\right)  $, $\mathbf{b}%
_{3}=-\mathbf{b}_{6}=a\left(  -\frac{\sqrt{3}}{2},\frac{3}{2}\right)  $. Here,
$\alpha_{s=1,3}$ or $\gamma_{s=1,2,3}$ is the gauge flux, and $\sigma_{x}$,
$\sigma_{y}$, and $\sigma_{z}$\ are the Pauli matrices for spin. $\delta$
denotes an on-site energy for sublattice A and B, and $h$ is the Zeeman
splitting. For simplicity, we choose $t=1$ as the energy unit and the distance
$a$ between the nearest sites as the length unit throughout this paper.

\begin{figure}[ptb]
\includegraphics[width=1.0\linewidth]{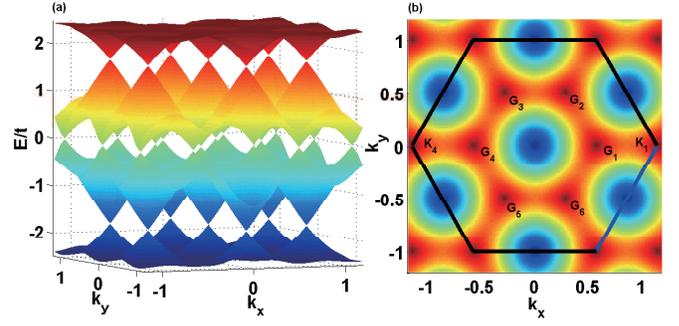}\caption{(Color online) (a)
Band structures for the Hamiltonian $H_{0}$ with $\alpha_{s=1,3}=\frac{\pi}%
{2}$. (b) Constant energy surfaces corresponding to the second band. The Dirac
points for the middle two bands lie at the $\mathbf{K}_{s=1,\cdots,6}$ points
and $\mathbf{G}_{s=1,\cdots,6}$ points in the first Brillouin zone (BZ). Here,
we only mark $\mathbf{K}_{1}$ and $\mathbf{K}_{4}$ and choose $\frac{2\pi}{3}$
as the unit of $k_{x}$ and $k_{y}$.}%
\end{figure}

In the momentum space, Hamiltonian (\ref{E0}) can be written in the basis
$\psi_{\mathbf{k}}=(a_{\mathbf{k}\uparrow},b_{\mathbf{k}\uparrow
},a_{\mathbf{k}\downarrow},b_{\mathbf{k}\downarrow})^{T}$:%
\begin{equation}
H=\sum_{\mathbf{k}}\psi_{\mathbf{k}}^{\dag}[\mathcal{H}_{0}\left(
\mathbf{k}\right)  +\mathcal{H}_{1}\left(  \mathbf{k}\right)  ]\psi
_{\mathbf{k}}. \label{Ham_m}%
\end{equation}
with%
\begin{align}
\mathcal{H}_{0}\left(  \mathbf{k}\right)   &  =\left(
\begin{array}
[c]{cccc}%
0 & -tP_{0}^{\ast}\left(  \mathbf{k}\right)  & 0 & -tP_{2}^{\ast}\left(
\mathbf{k}\right) \\
-tP_{0}\left(  \mathbf{k}\right)  & 0 & -tP_{1}\left(  \mathbf{k}\right)  &
0\\
0 & -tP_{1}^{\ast}\left(  \mathbf{k}\right)  & 0 & -tP_{0}^{\ast}\left(
\mathbf{k}\right) \\
-tP_{2}\left(  \mathbf{k}\right)  & 0 & -tP_{0}\left(  \mathbf{k}\right)  & 0
\end{array}
\right)  ,\label{E4}\\
\mathcal{H}_{1}\left(  \mathbf{k}\right)   &  =-\left(  t_{1}P_{3}\left(
\mathbf{k}\right)  -h\right)  \left(  \sigma_{z}\otimes \tau_{0}\right)
+\delta \left(  \sigma_{0}\otimes \tau_{z}\right)  , \label{E41}%
\end{align}
and%

\begin{align*}
P_{0}\left(  \mathbf{k}\right)   &  =e^{-i\mathbf{k\cdot d}_{2}},\\
P_{1}\left(  \mathbf{k}\right)   &  =ie^{-i\mathbf{k\cdot d}_{1}%
}+e^{-i\mathbf{k\cdot d}_{3}},\\
P_{2}\left(  \mathbf{k}\right)   &  =ie^{-i\mathbf{k\cdot d}_{1}%
}-e^{-i\mathbf{k\cdot d}_{3}},\\
P_{3}\left(  \mathbf{k}\right)   &  =2\sum_{s=1,2,3}\sin \left(  \mathbf{k\cdot
b}_{s}\right)  .
\end{align*}
Here, we set $\alpha_{s=1,3}=\gamma_{s=1,2,3}=\frac{\pi}{2}$. $\tau_{z}$ is
the Pauli matrix to span two sublattices. $\tau_{0}$ and $\sigma_{0}$ are
$2\mathtt{\times}2$ unit matrices. The energy spectrum for the Hamiltonian
$\mathcal{H}_{0}(\mathbf{k})$ is shown in Fig. 1 (a), where one sees that the
whole energy spectrum displays a \textquotedblleft
particle-hole\textquotedblright \ symmetry with respect to the Fermi energy
$\epsilon_{F}\mathtt{=}0$ and any two adjacent touching bands exhibit
Dirac-type energy spectra. The lower (upper) two bands touch at six
$\mathbf{M}$ points in the first Brillouin zone. The middle two bands touch at
six $\mathbf{K}$ points, only two of which are independent, with
$\mathbf{K}_{1,4}=\frac{2\pi}{3}\left(  \pm \frac{2\sqrt{3}}{3},0\right)  $,
$\mathbf{K}_{2,3,5,6}=\frac{2\pi}{3}\left(  \pm \frac{\sqrt{3}}{3},\pm1\right)
$ and $\mathbf{G}$-points with $\mathbf{G}_{1,4}=\frac{2\pi}{3}\left(
\pm \frac{\sqrt{3}}{3},0\right)  $, $\mathbf{G}_{2,3,5,6}=\frac{2\pi}{3}\left(
\pm \frac{\sqrt{3}}{6},\pm \frac{1}{2}\right)  $ in the first Brillouin zone
[shown in Fig. 1(b)]. Here, we choose $\frac{2\pi}{3}$ as the unit of $k_{x}$
and $k_{y}$ in all of the figures in this paper.

\section{Quantum Hall effects induced by perturbations}

To open a gap at each Dirac point, $\mathcal{H}_{1}\left(  \mathbf{k}\right)
$ in Eq. (\ref{Ham_m}) is turned on. The scheme for simulation of the gauge
field with $\gamma_{s=1,2,3}=\frac{\pi}{2}$ is presented in the Appendix. It
is straightforward to check that nonzero $t_{1}$, $\delta$, and $h$ can
independently open gaps at the eight inequivalent Dirac points. However, gaps
from different perturbations may have different topological features, and the
competitions or cooperations between these perturbations determine the overall
topological properties of the system. In order to explicitly show the
connections between the topological feature and the amplitudes of
perturbations, we expand the Hamiltonian in Eqs. (\ref{E4}) and (\ref{E41})
around each Dirac point and obtain the relevant low-energy Hamiltonian as follows:%

\begin{equation}
\mathcal{H}^{(m)}\left(  \mathbf{k}\right)  =d_{x}^{(m)}(\mathbf{k})\eta
_{x}+d_{y}^{(m)}(\mathbf{k})\eta_{y}+d_{z}^{(m)}(\mathbf{k})\eta
_{z}\label{Heff}%
\end{equation}
Here, $\eta_{x}$, $\eta_{y}$ and $\eta_{z}$\ are the Pauli matrices in the
effective band basis. $m$ labels the Dirac point, and $\mathbf{k}$ is measured
from the Dirac points. We summarize three components $(d_{x}^{(m)}%
(\mathbf{k}),d_{y}^{(m)}(\mathbf{k}),d_{z}^{(m)}(\mathbf{k}))$ of
$\mathbf{\hat{d}}^{(m)}(\mathbf{k})$ for all the Dirac points in Table
\ref{Dirac}.\begin{table}[pt]
\caption{Coefficients of the Dirac Hamiltonian at eight Dirac points.}
\begin{tabular}
[c]{llll}\hline \hline
& $d_{x}(\mathbf{k})$ & $d_{y}(\mathbf{k})$ & $d_{z}(\mathbf{k})$\\ \hline
$\mathbf{K}_{1}/\mathbf{G}_{4}$ & $-\frac{3tk_{x}}{\sqrt{6}}$ & $\frac
{3tk_{y}}{\sqrt{6}}$ & $\left(  \delta-t_{1}-\frac{\sqrt{3}h}{3}\right)  $\\
$\mathbf{K}_{4}/\mathbf{G}_{1}$ & $\frac{3tk_{x}}{\sqrt{6}}$ & $\frac{3tk_{y}%
}{\sqrt{6}}$ & $\left(  \delta-t_{1}+\frac{\sqrt{3}h}{3}\right)  $\\
$\mathbf{G}_{2}/\mathbf{G}_{5}$ & $\mp \frac{\sqrt{6}t\left(  k_{x}-\sqrt
{3}k_{y}\right)  }{4}$ & $\frac{\sqrt{6}t\left(  \sqrt{3}k_{x}+k_{y}\right)
}{4}$ & $\left(  \delta-3t_{1}\pm \frac{\sqrt{3}h}{3}\right)  $\\
$\mathbf{G}_{3}/\mathbf{G}_{6}$ & $\pm \frac{\sqrt{6}t\left(  k_{x}+\sqrt
{3}k_{y}\right)  }{4}$ & $\frac{\sqrt{6}t\left(  -\sqrt{3}k_{x}+k_{y}\right)
}{4}$ & $\left(  \delta+t_{1}\mp \frac{\sqrt{3}h}{3}\right)  $\\ \hline \hline
\end{tabular}
\label{Dirac}%
\end{table}The valley Chern number of the lower band around Dirac point $m$
can be calculated by \cite{Stic}%
\begin{equation}
C_{m}=\frac{1}{2}\text{sgn}(\mathbf{\partial}_{k_{x}}\mathbf{\hat{d}}%
^{(m)}(\mathbf{k})\times \mathbf{\partial}_{k_{y}}\mathbf{\hat{d}}%
^{(m)}(\mathbf{k}))_{z}\text{sgn}(d_{z}^{(m)}),\label{E15}%
\end{equation}
from which, we obtain%

\begin{align}
C_{\mathbf{K_{1}/G_{4}}}  &  =-\frac{1}{2}\text{sgn}\left(  \delta-t_{1}%
-\frac{\sqrt{3}}{3}h\right)  ,\label{c1}\\
C_{\mathbf{K_{4}/G_{1}}}  &  =\frac{1}{2}\text{sgn}\left(  \delta-t_{1}%
+\frac{\sqrt{3}}{3}h\right)  ,\label{c2}\\
C_{\mathbf{G_{2}/G_{5}}}  &  =\mp \frac{1}{2}\text{sgn}\left(  \delta-3t_{1}%
\pm \frac{\sqrt{3}}{3}h\right)  ,\label{c3}\\
C_{\mathbf{G_{3}/G_{6}}}  &  =\pm \frac{1}{2}\text{sgn}\left(  \delta+t_{1}%
\mp \frac{\sqrt{3}}{3}h\right)  . \label{c4}%
\end{align}

Having obtained the expressions for the valley Chern numbers, we now discuss
the influences of the perturbations on the topological properties of the
system. Generally speaking, the topological properties are the global features
of the system, and some physical quantities have simple corresponding
relations to the total Chern numbers, such as the Hall conductance
$\sigma_{xy}^{H}=-\frac{e^{2}}{h}\sum_{m}C_{m}$ in the electron system, and
the mass conductance $\sigma_{xy}^{M}\varpropto \sum_{m}C_{m}$ in the ultracold
atom system, with $\delta C_{m}$ denoting the change in Chern number with the
applied perturbations. For the two-Dirac-point system, $C_{m}$ can only take
thevvalue $0$ or $\pm1$, and the bulk-boundary correspondence is simple. For
the multi-Dirac-point system here, $C_{m}$ can take a series of values, and
the bulk-edge correspondence can represent abundant features due to the
extended internal valley degrees of freedom. Furthermore, edge states can show
different features for different boundaries.

In order to present an explicit picture of the bulk-boundary correspondence in
the multi-Dirac point system, we calculate the spectra of Hamiltonian
(\ref{E0}) with zigzag (armchair) boundaries along the $y(x)$ direction and
periodic boundaries along the $x(y)$ direction. The results are shown in Fig.
2 and 3. For a ribbon with zigzag boundaries, by comparing the magnitudes of
the gaps opened by $t_{1}$, $\delta$, and $h$ at each Dirac point, we can
identify that the points $\left(  \mathbf{G_{3},G_{5}}\right)  $ in the bulk
Brillouin zone are projected to the same momentum $k_{x}=-\frac{\sqrt{3}\pi
}{9}$ in Fig. 2. Similarly, the points $\left(  \mathbf{G_{2},G_{6}}\right)
$, $\left(  \mathbf{G_{1},K_{4}}\right)  $, and $\left(  \mathbf{K_{1},G_{4}%
}\right)  $ are projected to the momenta $k_{x}=\frac{\sqrt{3}\pi}{9}$,
$\frac{2\sqrt{3}\pi}{9}$, and $\frac{4\sqrt{3}\pi}{9}$, respectively. The same
conclusions can also be drawn from Fig. 1(b). Thus, it is convenient to divide
the eight Dirac points into two groups with each group involving four points.
Explicitly, group I includes points \{$\mathbf{G_{3}}$,\textbf{ }%
$\mathbf{G_{2}}$,\textbf{ }$\mathbf{G_{5}}$, $\mathbf{G_{6}}$\}, and group II
includes points \{$\mathbf{G_{1}}$,\textbf{ }$\mathbf{K_{1}}$,\textbf{
}$\mathbf{K_{4}}$, $\mathbf{G_{4}}$\}. See Fig. 2(a) for details. Furthermore,
each group can be divided into two subgroups and each subgroup includes two
Dirac points. Namely, two subgroups in group I are \{$\mathbf{G_{3},G_{2}}$\}
and\{$\mathbf{G_{5},G_{6}}$\}, while two subgroups in group II are
\{$\mathbf{G_{1},K_{1}}$\} and\{$\mathbf{K_{4},G_{4}}$\}. The topological
properties of Dirac points can be characterized by the valley Chern numbers
$C_{\mathbf{G_{i}}}$ and $C_{\mathbf{K_{i}}}$. In order to guarantee the
bulk-boundary correspondence, we define the joint valley Chern numbers
$C_{32}$, $C_{56}$, $C_{11}$, and $C_{44}$, in which $C_{32}=C_{\mathbf{G_{3}%
}}+C_{\mathbf{G_{2}}}$. Likewise, $C_{56}$, $C_{11}$, and $C_{44}$ have
similar definitions. For a ribbon with armchair boundaries, with a similar
strategy, we find that groups I and II are the same as in the case of a ribbon
with zigzag boundaries. However, we find from Fig. 3 (a) that the four Dirac
points in each group mix together. We show that double-pairing and
quadruple-pairing of Dirac points under different boundary conditions strongly
influence the transport behaviors along different types of edges. The
information on the edge states can be extracted from the distribution of the
number and spin density of the atoms along the edges. Define the number and
spin density of the atoms as follows,%

\begin{align}
n_{\alpha}^{(n)}(i) &  =\langle g|\mathbf{\alpha}_{n}^{\dag}(i)\mathbf{\sigma
}_{0}\mathbf{\alpha}_{n}(i)|g\rangle,\label{density_num}\\
s_{\alpha,\tau}^{(n)}(i) &  =\frac{1}{2}\langle g|\mathbf{\alpha}_{n}^{\dag
}(i)\mathbf{\sigma}_{\tau}\mathbf{\alpha}_{n}(i)|g\rangle.\label{density_spin}%
\end{align}
Here, $n_{\alpha}^{(n)}(i)$ and $s_{\alpha,\tau}^{(n)}(i)$ denote the number
and spin of atoms at site $i$ in the $n$th edge state. $\alpha=a,b$ marks two
sublattices and $\tau=x,y,z$ labels three components of spin. $\mathbf{\alpha
}_{n}(i)=[\alpha_{n,\uparrow}(i),\alpha_{n,\downarrow}(i)]^{\text{T}}%
$.\begin{figure}[ptb]
\includegraphics[width=1.0\linewidth]{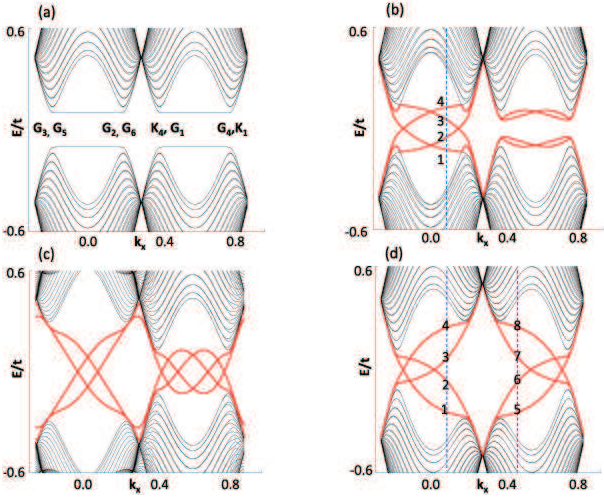}\caption{(Color online) Band
structures of the Hamiltonian (\ref{E0}) in the ribbon geometry with zigzag
edges and 40 sites in the $y$ direction, corresponding to (a) $\left(
\delta,t_{1},h\right)  =\left(  0.1,0,0\right)  $, (b) $\left(
0.1,0.05,0\right)  $, (c) $\left(  0.1,0.2,0\right)  $, (d) $\left(
0.1,0,0.3\right)  $ Here, $k_{x}$ is measured in units of $2\pi/3$.}%
\end{figure}

\begin{figure}[ptb]
\includegraphics[width=1.0\linewidth]{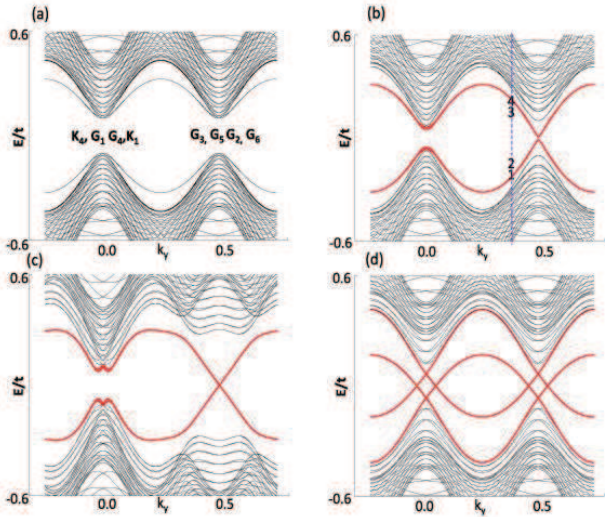}\caption{ (Color online) Band
structures of Hamiltonian (\ref{E0}) in the ribbon geometry with armchair
edges and 60 sites in the $x$ direction, corresponding to (a) $\left(
\delta,t_{1},h\right)  =\left(  0.1,0,0\right)  $, (b) $\left(
0.1,0.05,0\right)  $, (c) $\left(  0.1,0.2,0\right)  $, and (d) $\left(
0.1,0,0.3\right)  $. Here, $k_{y}$ is measured in units of $2\pi/3$.}%
\end{figure}

Now, we discuss the evolution of the bulk-boundary correspondence by turning
on different perturbations. For a ribbon with zigzag boundaries, starting with
the nonzero $\delta$, the edge spectrum is shown in Fig. 2(a), from which no
gapless edge states emerge. Correspondingly, the joint valley Chern number
$C_{32}=C_{56}=C_{11}=C_{44}=0$. The system lies in a trivial state with the
total Chern number $C_{tot}=C_{32}+C_{56}+C_{11}+C_{44}$ in the half-filling
case. When $t_{1}$ is turned on, one can observe from Eq. (\ref{c1}-\ref{c4})
that only $C_{\mathbf{G_{2}}}$ and $C_{\mathbf{G_{5}}}$ change signs when
$t_{1}\in \lbrack \delta/3,\delta]$. Then, we have $C_{32}=C_{\mathbf{G_{2}}%
}|_{t_{1}>\delta/3}-C_{\mathbf{G_{2}}}|_{t_{1}<\delta/3}=+1$, and
$C_{56}=C_{\mathbf{G_{5}}}|_{t_{1}>\delta/3}-C_{\mathbf{G_{5}}}|_{t_{1}%
<\delta/3}=-1$. Meanwhile, two pairs of gapless edge states emerge [Fig.
2(b)], and each pair of edge states connects each pair of Dirac points
\{$\mathbf{G_{3},G_{2}}$\} and \{$\mathbf{G_{5},G_{6}}$\}, respectively. To
show the features of edge states, we plot the distribution of $n_{\alpha
}^{(n)}(i)$ and $s_{\alpha,\tau}^{(n)}(i)$ in Figs. 4(a1)-4(a4), and we can
determine the propagation direction of the $n$th edge state through the
velocity $\upsilon=\partial E_{n}\left(  k_{x}\right)  /\partial k_{x}$. The
opposite propagation directions correspond to the opposite signs of the valley
Chern numbers $C_{32}$ and $C_{56}$. The consistency between the valley Chern
number $C_{32}$, $C_{56}$ and the gapless edge states satisfies the
bulk-boundary correspondence. The transport pictures of the edge states are
schematically illustrated in Fig. 4(c), where the edge states for $C_{32}=+1$
move along the clockwise direction and the edge states for $C_{56}=-1$ move
along the counterclockwise direction. Furthermore, from Figs. 4(a2)-4(a4), we
find that $s_{a,z}^{(1)}(i)$ and $s_{a,z}^{(2)}(i)$ exponentially decay as the
distance to the boundary with a positive-negative oscillation, while
$s_{a,x/y}^{(1)}(i)$ and $s_{a,x/y}^{(2)}(i)$ exponentially and monotonously
decay as the distance to the boundary. The $s_{b,\tau}^{(n)}(i)$ have similar
behaviors. Thus, we clarify that the spin polarizations of edge states almost
lie in the $a$-$b$ plane, edge states with opposite spin polarizations
propagate at opposite velocities at each boundary of ribbon (see Fig. 4(c)),
and states with $t_{1}\in \lbrack \delta/3,\delta]$ are QSH states. Compared
with the standard QSH state proposed in the Kane-Mele model \cite{Kane} which
involves four bands, the number of effective bands in our model is two, but
the increased internal valley degrees of freedom compensate the minimum
four-band requirement to induce QSH states. In other multi-Dirac-point models
\cite{Eza,Bena}, similar QSH states might exist if the gaps can been opened by
some special perturbations. Of more practical significance, Ref. \cite{An} has
studied the transport properties of a silicene model with two kinds of Rashba
spin-orbit couplings \cite{Yao}, where the QSH phase has similar features to
our model. When $t_{1}$ is tuned to be larger than $\delta$, we find that
$C_{\mathbf{K_{1}}},C_{\mathbf{G_{4}}},C_{\mathbf{K_{2}}}$, and
$C_{\mathbf{G_{1}}}$ in group \textrm{II} change their signs, and the joint
valley Chern number $C_{11}=C_{44}=0$. Even though two pairs of edge states
emerge [see Fig. 2(c)], they can be adiabatically tuned into the bulk. The
topological properties are similar to the case with $t_{1}\in \lbrack
\delta/3,\delta]$. For a ribbon with armchair boundaries, the joint valley
Chern numbers $C_{32}$ and $C_{56}$ merge into a single one, $C_{3}%
+C_{5}+C_{2}+C_{6}$, as shown in Fig. 3(a). They become indistinguishable, and
the joint valley Chern numbers can not be well defined. Correspondingly, the
edge states also mix together in Figs. 4(b1)-4(b4). These can be called
quasiedge states, because they cannot induce the net spin current but only
accumulate atoms at the armchair edges. We schematically show the features of
quasi-edge states in Fig. 4 (d). Actually, the different behaviors of the edge
states between zigzag and armchair boundaries can be understood as follows.
Zigzag boundaries have an A-sublattice terminal at one boundary and a
B-sublattice terminal at the other boundary, while armchair boundaries have
both A- and B-sublattice terminals at both boundaries. Figures 4(a1)--4(a4)
show that the chirality of the edge state can only be well defined when the
edge state is localized on one kind of sublattice. This is the physical reason
for the difference between ribbons with zigzag and ribbons with armchair
boundaries. These features indicate that the transport behaviors of edge
states from different types of edges can be very different in
multi-Dirac-point systems even though the globally structures are
characterized by a single topological number. \begin{figure}[ptb]
\includegraphics[width=1.0\linewidth]{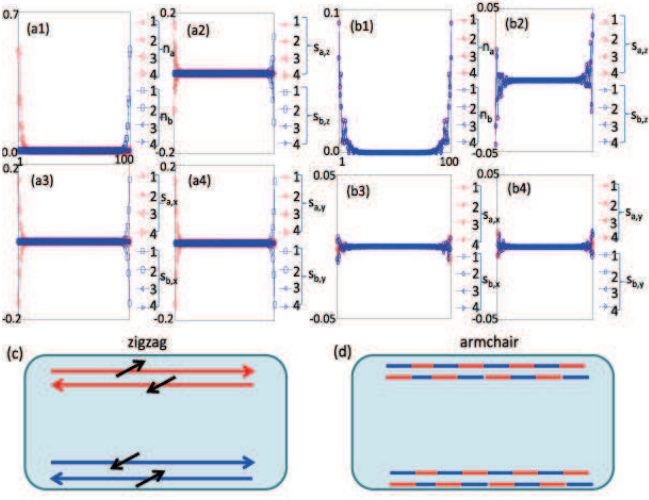}\caption{(Color online)
(a1)-(a4) [(b1)-(b4)] Distributions of the atom density $n_{\alpha}$ and three
spin components ($s_{\alpha x},s_{\alpha y},s_{\alpha z}$) along a zigzag
[armchair] ribbon with 100 lattice sites for four edge states in Fig. 2(b)
[Fig. 3(b)], labeled $1$, $2$, $3$, and $4$. Here, $\alpha=a,b$ denote two A
and B sublattices. (c) Schematic of the movement of atoms in edge states shown
in Fig. 2(b). The spin-$s_{z}$ component is nearly suppressed, while
spin-$s_{x}$ and -$s_{y}$ components have high weights. (d) Schematic of the
accumulations of atoms in edge states corresponding to Fig. 3(b). The spin
polarization is strongly suppressed, and no net-current can be driven. }%
\end{figure}

Now, we turn to the effect of Zeeman splitting. Suppose that the system lies
in a trivial state initially with $\delta \neq0$ (Fig. 2(a)). Turning to $h$,
when $h>\sqrt{3}\delta$, $C_{\mathbf{K_{1}}}$, $C_{\mathbf{G_{3}}}$,
$C_{\mathbf{G_{4}}}$, and $C_{\mathbf{G_{5}}}$ change their signs. For a
ribbon with zigzag boundaries, straight forward calculations show that the
joint valley Chern numbers $C_{32/56}=C_{\mathbf{G_{3/5}}}|_{h>\sqrt{3}\delta
}-C_{\mathbf{G_{3/5}}}|_{h<\sqrt{3}\delta}=-1$ and $C_{44/11}=C_{\mathbf{G_{4}%
/K}_{1}}|_{h>\sqrt{3}\delta}-C_{\mathbf{G_{4}/K}_{1}}|_{h<\sqrt{3}\delta}=+1$,
and there are four pairs of gapless edge states, shown in Fig. 2(d). Further
analysis identifies that the two pairs of edge states connecting Dirac points
in group \textrm{I} with $C_{32}=C_{56}=-1$ propagate along the
counterclockwise direction and the edge states connecting Dirac points in
group \textrm{II} with $C_{44}=C_{11}=+1$ propagate along the clockwise
direction, as shown in Figs. 5(c)-5(d). In Figs. 5(a1)-5(b4), we see that
$s_{a,x}^{(n)}(i)$ and $s_{a,y}^{(n)}(i)$ are 0 while $s_{a,z}^{(n)}(i)$ is
nonzero for $n=1$...$8$, and the relevant spin polarization directions of the
edge-state are also shown in Figs. 5(c)-5(d). Since the total Chern number is
equal to 0, one can not distinguish this topological phase by measuring the
Hall conductance. Figure 2(d) shows that the edge states are separated into
two groups according to valleys with respect to $k_{x}=\frac{\sqrt{3}\pi}{6}$.
Thus we call these quantum anomalous valley Hall states. If other
perturbations or intervalley scatterings are introduced to break the
topological structures in group I or II, the quantum anomalous valley Hall
states become measurable. For a ribbon with armchair boundaries, the joint
valley Chern number defined in the zigzag case shows the same behaviors, i.e.,
$C_{32}=C_{56}=-1$, and $C_{44}=C_{11}=+1$. Therefore, the quadruple-pairing
of them does not induce destructive mixtures and the spectra of the edge
states are shown in Fig. 3(d); they show features similar to those in Fig.
2(d), due to having the same joint valley Chern numbers $C_{32}=C_{56}=-1$,
and $C_{44}=C_{11}=+1$.

\begin{figure}[ptb]
\includegraphics[width=1.0\linewidth]{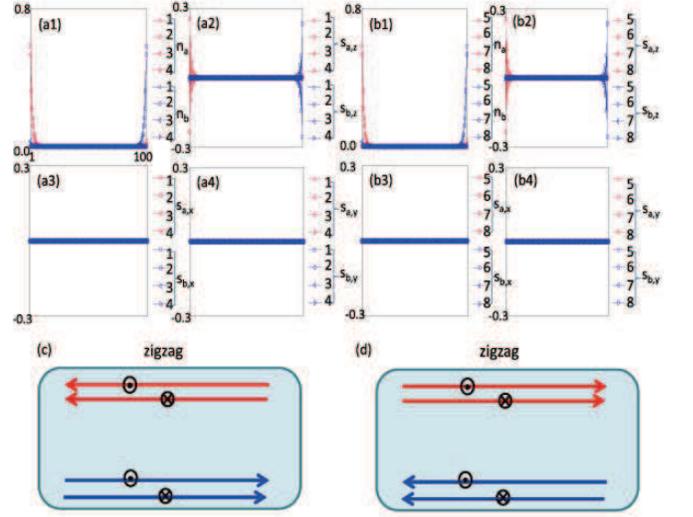}\caption{(Color online)
(a1)-(b4) Distributions of the atom density $n_{\alpha}$ and three spin
components ($s_{\alpha x},s_{\alpha y},s_{\alpha z}$) along a zigzag ribbon
with 100 lattice sites for eight edge states in Fig. 2(d), labeled $1$ to $8$,
respectively. Here, $\alpha=a,b$ denote two A and B sublattices. (c) Schematic
of the movement of atoms in edge states in Fig. 2(d) labeled 1 to 4. (d)
Schematic of the movement of atoms in edge states in Fig. 2(d) labeled 5 to 8.
}%
\end{figure}

\begin{figure}[ptb]
\includegraphics[width=1.0\linewidth]{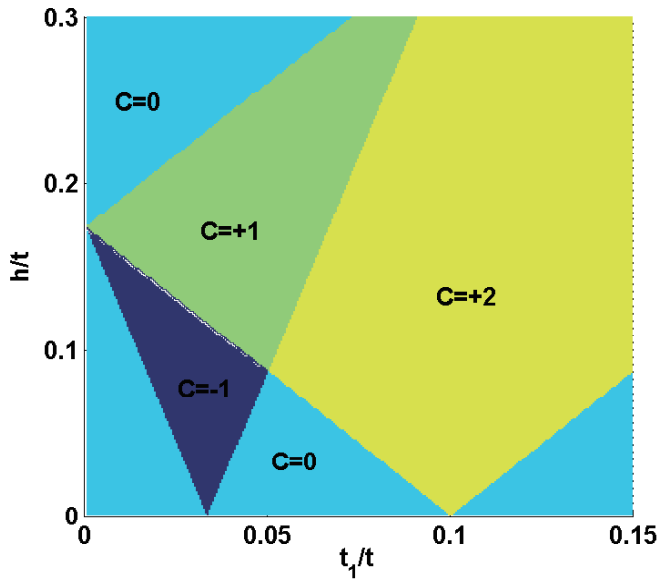}\caption{(Color online) Phase
diagram in the $t_{1}-h$ plane. Here, we set $\delta=0.1t.$}%
\end{figure}

Having obtained the individual picture of nonzero $t_{1}$ and $h$, we turn to
nonzero $t_{1}$ and nonzero $h$ combined. Then the Hall conductance or mass
conductance becomes parameter dependent and is determined by the total Chern
number. We plot the phase diagram about $t_{1}/t$ and $h/t$ with $\delta=0.1$
in Fig. 6, from which shows that the Chern number can be tchanged from 0 to 2
by $t_{1}$ and $h$ in this multi-Dirac-point system. To clearly see the
correspondence between the Chern number and the edge states of zigzag and
armchair boundaries, we start with the parameters $(\delta$, $t_{1}$,
$h)=(0.1$, $0.0$, $0.0)$ initially, and we have $C_{32}=C_{56}=C_{11}%
=C_{44}=0$. When $(\delta$, $t_{1}$, $h)=(0.1$, $0.03$, $0.11)$, only
$C_{\mathbf{G_{5}}}$ changes its sign and we have $C_{56}=C_{\mathbf{G_{5}}%
}|_{\delta<3t_{1}+\frac{\sqrt{3}}{3}h}-C_{\mathbf{G_{5}}}|_{\delta
>3t_{1}+\frac{\sqrt{3}}{3}h}=-1$. Thus, the system is in the QAH state with
$C_{tot}=C_{56}=-1$. A pair of edge states emerges [see Figs. 7(a1) and 7(b1)
for the cases of zigzag boundaries and armchair boundaries]. When $(\delta$,
$t_{1}$, $h)=(0.1$, $0.03$, $0.18)$, then $C_{\mathbf{K_{1}}}$ and
$C_{\mathbf{G_{4}}}$ also change signs. We obtain QAH states with
$C_{44}=C_{11}=+1$ and the Chern number of the whole system $C=C_{56}%
+C_{44}+C_{11}=-1+1+1=+1$. The corresponding three edge states are shown in
Figs. 7(a2) and 7(b2); two pairs have positive chirality and one pair has
negative chirality. Furthermore, when $(\delta$, $t_{1}$, $h)=(0.1$, $0.1$,
$0.07)$, $C_{\mathbf{G_{2}}}$ changes sign and we obtain QAH states with
$C=+2$. In Fig. 7(a3), we see that the edge states corresponding to
$C_{56}=-1$ and $C_{23}=1$ couple with each other and open a gap to destroy
nontrivial features, while the edge states corresponding to $C_{11}=1$ and
$C_{44}=1$ show nontrivial features, and retain the overall Chern number
$C=+2$. Note that even the leftward two valleys in Fig. 3(b) have Chern
numbers $C_{56}=-1$ and $C_{23}=1$, but no gap is opened between them in
comparison with the edge states in Fig. 7(a3). The reason is that the
time-reversal symmetry is conserved in Fig. 3(b), and this symmetry protects
the QSH state. On the other hand, the time-reversal symmetry is broken in Fig.
7(a3), and the two edge states can couple with each other and a band gap is
opened. In Fig. 7(b3), it is straightforward to check that the two pairs of
edge states at the quadruple-pairing Dirac points \{$\mathbf{G_{3}}$,\textbf{
}$\mathbf{G_{2}}$,\textbf{ }$\mathbf{G_{5}}$, $\mathbf{G_{6}}$\} give the
Chern number $C=C_{3}+C_{5}+C_{2}+C_{6}=0$, and the overall Chern number
$C=+2$ is protected by another quadruple-pairing of Dirac points,
\{$\mathbf{G_{1}}$,\textbf{ }$\mathbf{K_{1}}$,\textbf{ }$\mathbf{K_{4}}$,
$\mathbf{G_{4}}$\}, with Chern number $C=1+1=+2$. Note that in Fig. 7 (b3),
the gapless edge states in the right valley are fake and not stable, and
increasing Zeeman splitting can break these fake gapless edge states. Thus,
the bulk-boundary correspondence is globally robust, however, the edge states
represent different features for different boundaries.

\begin{figure}[ptb]
\includegraphics[width=1.0\linewidth]{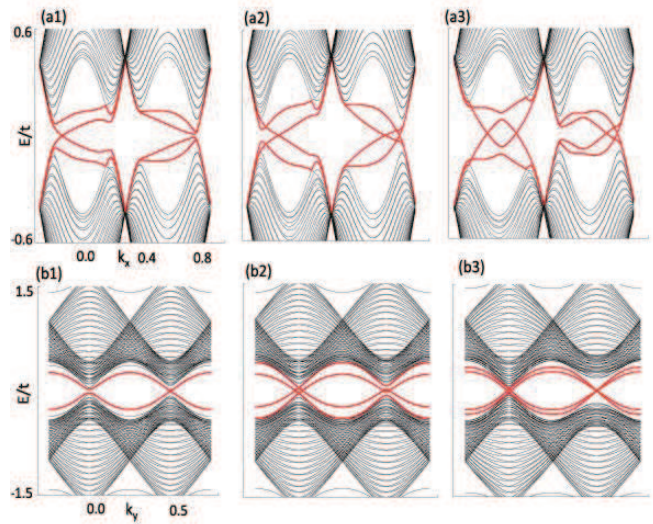}\caption{ (Color online)
(a1)-(a3) [(b1)-(b3)] Band structures of Hamiltonian (\ref{E0}) for the zigzag
[armchair] ribbon geometry with 40 [60] sites in the $y$ [$x$] direction with
parameters $\left(  \delta,t_{1},h\right)  =\left(  0.1,0.03,0.11\right)  $,
$\left(  0.1,0.03,0.18\right)  $, and $\left(  0.1,0.1,0.07\right)  $,
respectively. These three cases show the edge states of the QAH effect with
the Chern number of the whole system $C$=$-1$, $+1$, and $+2$. Here, $k_{x}$,
and $k_{y}$ are measured in units of $2\pi/3$.}%
\end{figure}

\section{Discussions and summary}

In principle, the largest Chern number for this eight-Dirac-point system is 4
\cite{Stic}. However, the phase diagram in Fig. 6 shows that the largest Chern
number is 2. For the present three kinds of perturbations, we cannot obtain a
$C=4$ phase. If we introduce more perturbations to increase the numbers of
parameters in Eqs. (8)--(11), we believe that the $C=4$ phase can be achieved.
For example, the Chern numbers of $K_{1}$, $G_{4}$, $G_{2}$, and $G_{6}$
change signs, while the other four do not change signs. Thus, we can obtain
the $C=4$ phase. We focus on three kinds of perturbations in the present paper
and leave the realizations of the higher Chern number phases for further studies.

In the electron system, the Hall conductance is well defined as the response
coefficient of the electron current about the gradient of the electric
potential, i.e., the electric field intensity. Similarly, in the ultracold
atom system, the Hall conductance corresponds to the response coefficient of
the atom mass current about the gradient of the optical trapping potential.
Thus, the non-zero Hall conductance gives the experimental signatures of the
mass current and accumulations of atoms at the boundary of the optical lattice
when the trapping potential is modulated.

In summary, we study quantum Hall effects in a non-Abelian honeycomb optical
lattice which is a multi-Dirac-point system. The Hall conductance has
different values with the change of relative strengths of several
perturbations. We find that gauge-field-dressed NNN hopping can induce the QSH
effect, and a Zeeman field can induce the so-called quantum anomalous valley
Hall effect, which includes two copies of quantum Hall states with opposite
Chern numbers and counter-propagating edge states. The coexistence of both
perturbations gives the nonzero Hall conductance which is characterized by
different Chern numbers. Our study shows the power of extended internal valley
degrees of freedom in driving abundant quantum Hall effects in a non-Abelian
honeycomb optical lattice.

\section{Acknowledgments}

G. Liu was supported by NSF of China (No.11247011), This work was supported by
the NSFC under grants Nos. 11434015, 61227902, 61378017, 11374354, 11174360,
11575051, 51201057, NKBRSFC under grants Nos. 2011CB921502, 2012CB821305,
SKLQOQOD under grants No. KF201403, SPRPCAS under grants No. XDB01020300 and
Hebei NSF (Grant No. A2012202022, A2015208024).

\appendix

\section{Simulating the gauge-field-dressed next-nearest-neighbor hopping}

\label{Appendix} In this section, we focus on how to simulate the
gauge-field-dressed NNN hopping, i.e., the first term in Eq. (\ref{E2}), in
our model in optical lattices. The first term in the Eq. (\ref{E2}) actually
corresponds to a specific spin-orbit coupling if we take $\gamma_{s=1,2,3}%
=\pm \frac{\pi}{2}$. It is well known that in the Haldane model \cite{Hald}, a
periodic vector potential applied to a two-dimensional honeycomb lattice with
zero net flux through each unit cell. The spacial magnetic flux density does
not change the nearest-neighbor hopping amplitude, but it really causes the
NNN hopping to have a chirality and breaks the time-reversal symmetry [see
Fig. 8(a)]. Subsequently, Kane and Mele extended this mechanism to the
time-reversal-invariant system, with up-spin and down-spin electrons having
opposite chirality when they are hopping between NNN sites \cite{Kane}. This
kind of spin-dependent effective magnetic field is the intrinsic spin-obit
coupling in the Kane-Mele model. In Fig. 8, arrows between NNN sites denote
the directions of positive phase hopping for down-spin electrons. The
parameters $\gamma_{s=1,3}=\frac{\pi}{2}$ and $\gamma_{2}=-\frac{\pi}{2}$
correspond to the spin-dependent effective magnetic field in Fig. 8(b) and
$\gamma_{s=1,2,3}=\frac{\pi}{2}$ correspond to Fig. 8(c). It is surprising
that these two kinds of effective magnetic fields can give similar results for
the system, and the latter corresponds to the results in the text. For
comparison, we also show the Haldane model or Kane-Mele model in Fig. 8(a).

The spacial periodic magnetic field in the Haldane model can be simulated by
use of the laser-induced-gauge-field method in an optical lattice \cite{Zhu3}.
With a similar method, we can design two other kinds of effective magnetic
field. To this end, we consider a cold-atomic system with each atom having an
$\Lambda$-type level configuration \cite{Liu}. The ground states $\left \vert
1\right \rangle $ and $\left \vert 2\right \rangle $ are coupled to the excited
state $\left \vert 3\right \rangle $ through a spatially varying standing-wave
laser field, with Rabi frequencies $\Omega_{p}$=$\Omega \sin \theta e^{-iS_{1}}$
and $\Omega_{c}$=$\Omega \cos \theta e^{-iS_{2}}$, respectively. With the
rotating-wave approximation, the laser-atom coupling Hamiltonian is given by
\begin{equation}
\hat{H}_{\text{int}}=-\frac{\hbar}{2}\left(
\begin{array}
[c]{ccc}%
0 & 0 & \Omega_{p}\\
0 & 0 & \Omega_{c}\\
\Omega_{p}^{\ast} & \Omega_{c}^{\ast} & -2\Delta
\end{array}
\right)  \label{HS6}%
\end{equation}
with eigenstates (dressing states)$\allowbreak \allowbreak \allowbreak$
\begin{align}
\left \vert \chi_{1}\right \rangle  &  \text{=}e^{-iS_{1}}\cos \theta \left \vert
1\right \rangle \text{\texttt{-}}e^{-iS_{2}}\sin \theta \left \vert 2\right \rangle
\\
\left \vert \chi_{2}\right \rangle  &  \text{=}\cos \varphi \sin \theta e^{-iS_{1}%
}\left \vert 1\right \rangle \text{+}\cos \varphi \cos \theta e^{-iS_{2}}\left \vert
2\right \rangle \mathtt{-}\sin \varphi \left \vert 3\right \rangle \\
\left \vert \chi_{3}\right \rangle  &  \text{=}\sin \varphi \sin \theta e^{-iS_{1}%
}\left \vert 1\right \rangle \mathtt{+}\sin \varphi \cos \theta e^{-iS_{2}%
}\left \vert 2\right \rangle \text{\texttt{+}}\cos \varphi \left \vert
3\right \rangle
\end{align}
and eigenvalues $\lambda_{n=1,2,3}$=$0$,$\frac{\hbar}{2}\left(  \allowbreak
\Delta \mathtt{\mp}\sqrt{\Delta^{2}\text{+}\Omega^{2}}\right)  $. Here, the
single-photon detuning $\Delta=\omega_{3}\mathtt{-}\omega_{1}\mathtt{-}%
\omega_{p}$, with $\omega_{3}$, $\omega_{1}$, and $\omega_{p}$ the intrinsic
frequency of atom states $\left \vert 3\right \rangle $, atom state $\left \vert
1\right \rangle $, and laser $\Omega_{p}$, respectively. In the new basis space
$\left \vert \chi \right \rangle $=$\left \{  \left \vert \chi_{1}\right \rangle
\text{, }\left \vert \chi_{2}\right \rangle \text{, }\left \vert \chi
_{3}\right \rangle \right \}  $, the primary atom Hamiltonian $\hat{H}$%
=$\frac{\mathbf{p}^{2}}{2M}$+$\hat{H}_{\text{int}}\left(  \mathbf{r}\right)
$+$\hat{V}\left(  \mathbf{r}\right)  $ can be rewritten as $H$=$\frac{1}%
{2M}\left(  -i\hbar \nabla \mathtt{-}\mathbf{A}\right)  ^{2}$+$V$ with $M$ the
atom mass and $\mathbf{A}$ and $V$ the matrix with matrix elements
$\mathbf{A}_{n,m}$=$i\hbar \left \langle \chi_{n}\left(  \mathbf{r}\right)
\mathtt{\mid}\nabla \chi_{m}\left(  \mathbf{r}\right)  \right \rangle $ and
$V_{n,m}$=$\lambda_{n}\left(  \mathbf{r}\right)  \delta_{n,m}$+$\left \langle
\chi_{n}\left(  \mathbf{r}\right)  \right \vert \hat{V}\left(  \mathbf{r}%
\right)  \left \vert \chi_{m}\left(  \mathbf{r}\right)  \right \rangle $,
respectively. One can see that in the new basis the atom can be considered as
moving in gauge potential $\mathbf{A}$, which corresponds to an effective
magnetic field $\mathbf{B}_{eff}$=$\left(  \mathbf{\nabla \times A}\right)
\mathtt{-}\frac{i}{\hbar}\left(  \mathbf{A\times A}\right)  $ \cite{Zhu3,Ruse}.

\begin{figure}[ptb]
\includegraphics[width=1.0\linewidth]{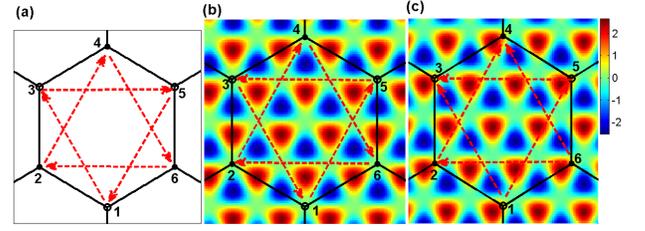}\caption{(Color online)
Illustration of the contours of the spin-dependent effective magnetic field.
Arrows between the next-nearest-neighbor sites denote the directions of
positive phase hopping for down-spin electrons. (a) The Haldane model; . (b)
the parameters $\gamma_{s=1,3}=\frac{\pi}{2}$, $\gamma_{2}=-\frac{\pi}{2}$,
and $\phi=\pi/2$; (c) the parameters $\gamma_{s=1,2,3}=\frac{\pi}{2}$ and
$\phi=2\pi/3$.}%
\end{figure}

We focus on the subspace spanned by the two lower eigenstates $\left \{
\left \vert \chi_{1}\right \rangle \text{, }\left \vert \chi_{2}\right \rangle
\right \}  $, which is redefined by $\left \vert \chi_{\uparrow}\right \rangle
\mathtt{\equiv}\left \vert \chi_{1}\right \rangle $ and $\left \vert
\chi_{\downarrow}\right \rangle \mathtt{\equiv}\left \vert \chi_{2}\right \rangle
$ in the spin language. This gives an effective spin-1/2 system. For the large
detuning ($\Delta \mathtt{\gg}\Omega$) case, both state $\left \vert
\chi_{\uparrow}\right \rangle $ and state $\left \vert \chi_{\downarrow
}\right \rangle $ are stable under atomic spontaneous emission from the initial
excite state $\left \vert 3\right \rangle $. Furthermore, we assume the
adiabatic condition, which requires that the off-diagonal elements of the
matrices $\mathbf{A}$ are smaller than the eigenenergy differences $\left \vert
\lambda_{m}\mathtt{-}\lambda_{n}\right \vert $ ($m$, $n$=$1$,$2$,$3$) of states
$\left \vert \chi_{m}\right \rangle $. Under this adiabatic condition and in the
basis space $\left \{  \left \vert \chi_{\uparrow}\right \rangle \text{,
}\left \vert \chi_{\downarrow}\right \rangle \right \}  $, the gauge potential
$\mathbf{A}$ becomes diagonal and has the form \cite{Zhu2}%
\begin{equation}
\mathbf{A=}\left(
\begin{array}
[c]{cc}%
\mathbf{A}_{\uparrow} & 0\\
0 & \mathbf{A}_{\downarrow}%
\end{array}
\right)
\end{equation}
with
\begin{equation}
\mathbf{A}_{\uparrow}=-\mathbf{A}_{\downarrow}=\hbar \left(  \nabla S_{1}%
\cos^{2}\theta \text{+}\nabla S_{2}\sin^{2}\theta \right)  .
\end{equation}
Here we neglect the correction to nearest-neighbor tunneling induced by the
change in potential $V\left(  \mathbf{r}\right)  $ thanks to the large
detuning approximation.

Consider the configuration of two opposite-traveling standing-wave laser beams
\cite{Zhu2,Zhu3}, which take Rabi frequencies $\Omega_{p}$=$\Omega \sin \left(
k_{2}y-\phi \right)  e^{-ik_{1}x}$ and $\Omega_{c}$=$\Omega \cos \left(
k_{2}y-\phi \right)  e^{ik_{1}x}$. The effective gauge potential is generated
as $\mathbf{A}_{\uparrow}$=$-\mathbf{A}_{\downarrow}$=$\hbar k_{1}\cos \left(
2k_{2}y-2\phi \right)  \mathbf{e}_{x}$. Here $k_{1}$=$k\sin \theta_{1}$, $k_{2}%
$=$k\cos \theta_{1}$, with $k$ the wave-vector number of the laser and
$\theta_{1}$ the angle between the wave vector and the $\mathbf{e}_{y}$ axis.
In order to obtain the special spin-orbit coupling in our model, the value of
$k_{2}$ should be selected appropriately. When the wave vector $k_{2}$ of the
laser beams satisfies $k_{2}=2l\pi$ with $l=1,2,3,\cdots$, the laser beams are
commensurate with the optical lattice. Furthermore, we can obtain a series of
Peierls phase factors which satisfy our model. For example, we take
$k_{2}a=2\pi$. The Peierls phase factors for the nearest neighbor hopping in
Fig. 8(b) are 0, and those for the NNN hopping in Fig. 8(b) are $\varphi
_{35}^{\alpha}$=$-\varphi_{62}^{\alpha}$=$\alpha \sqrt{3}k_{1}\cos \left(
2\phi \right)  $ and $\varphi_{13}^{\alpha}$=$\varphi_{51}^{\alpha}$%
=$\varphi_{24}^{\alpha}$=$\varphi_{46}^{\alpha}$=$0$, with $\alpha$=$\pm1$
representing the up- and down-spin. Considering the $C_{3}$ rotational
symmetry of the hopping phase factors in Fig. 8(b), we can rotate the vector
potential $\mathbf{A}$ by $\pm2\pi/3$ to obtain the other two vector
potentials. Therefore, the total effective vector potential and magnetic field
can be written as%
\begin{align}
\mathbf{A}_{eff}^{\alpha} &  \text{=}\alpha \hbar k_{1}\left[  \left(
\cos \left(  2k_{2}y-2\phi \right)  -\cos \left(  k_{2}y+2\phi \right)
\cos \left(  \sqrt{3}k_{2}x\right)  \right)  \mathbf{e}_{x}\right.  \nonumber \\
&  \left.  -\sqrt{3}\hbar k_{1}\sin \left(  k_{2}y+2\phi \right)  \sin \left(
\sqrt{3}k_{2}x\right)  \mathbf{e}_{y}\right]  ,\label{A1}%
\end{align}%
\begin{equation}
\mathbf{B}_{eff}^{\alpha}\text{=}\mathtt{-}\alpha4\pi \hbar k_{1}\left[
2\sin \left(  2\pi y+2\phi \right)  \cos \left(  \frac{6\pi}{\sqrt{3}}x\right)
-\sin \left(  4\pi y-2\phi \right)  \right]  \mathbf{e}_{z}.\label{B1}%
\end{equation}
The total accumulated phases for nearest-neighbor hopping are 0, and those for
the NNN hopping along the arrowed directions in Fig. 8(b) are%
\begin{equation}
\varphi_{13}^{\alpha}=\varphi_{35}^{\alpha}=\varphi_{51}^{\alpha}%
=\mathbf{-}\varphi_{24}^{\alpha}=-\varphi_{46}^{\alpha}=-\varphi_{62}^{\alpha
}=\alpha \sqrt{3}k_{1}\cos \left(  2\phi \right)
\end{equation}
The contours of the magnetic field with $\phi=\pi/2$ for down-spin are also
plotted in Fig. 8(b).

Comparing Fig. 8(b) with Fig. 8(c), we find that the phase factors have
different signs along the sites $\left(  1\rightarrow3\right)  $ direction and
the latter does not have the $C_{3}$ rotational symmetry. So, along this
direction, we select the Rabi frequencies $\Omega_{p}$=$\Omega \sin \left(
-\frac{\sqrt{3}}{2}k_{2}x-\frac{1}{2}k_{2}y-\phi \right)  e^{i\left(  -\frac
{1}{2}k_{1}x+\frac{\sqrt{3}}{2}k_{1}y\right)  }$ and $\Omega_{c}$=$\Omega
\cos \left(  -\frac{\sqrt{3}}{2}k_{2}x-\frac{1}{2}k_{2}y-\phi \right)
e^{-i\left(  -\frac{1}{2}k_{1}x+\frac{\sqrt{3}}{2}k_{1}y\right)  }$ and still
set $k_{2}a=2\pi$. The laser field along the $\left(  1\rightarrow3\right)  $
direction brings the non-zero phase factor $\varphi_{13}=-\varphi_{46}%
=-\alpha \sqrt{3}k_{1}a\cos \left(  2\phi \right)  $. Finally, the total
effective vector potential and magnetic field for the Fig. 8(c) can be written
as
\begin{align}
\mathbf{A}_{eff} &  =\hbar k_{1}\left[  \cos \left(  2k_{2}y-2\phi \right)
-\sin \left(  k_{2}y+2\phi \right)  \sin \left(  \sqrt{3}k_{2}x\right)  \right]
\mathbf{e}_{x}\nonumber \\
&  -\sqrt{3}\hbar k_{1}\cos \left(  k_{2}y+2\phi \right)  \cos \left(  \sqrt
{3}k_{2}x\right)  \mathbf{e}_{y}%
\end{align}
and%
\begin{align}
\mathbf{B}_{eff}^{\alpha} &  \text{=}\alpha4\hbar k_{1}\pi \left[  2\cos \left(
\frac{2\pi}{a}y+2\phi \right)  \sin \left(  \frac{2\sqrt{3}\pi}{a}x\right)
\right.  \nonumber \\
&  \left.  +2\sin \left(  \frac{4\pi}{a}y-2\phi \right)  \right]  \mathbf{e}_{z}%
\end{align}
The phase factors along the arrowed direction in Fig. 8(c) are
\begin{equation}
\varphi_{35}^{\alpha}=\varphi_{51}^{\alpha}=-\varphi_{13}^{\alpha}%
=\varphi_{46}^{\alpha}=\mathbf{-}\varphi_{24}^{\alpha}=-\varphi_{62}^{\alpha
}=\alpha \sqrt{3}k_{1}\cos \left(  2\phi \right)
\end{equation}
The contours of the magnetic field with $\phi=2\pi/3$ for down-spin are
plotted in Fig. 8(c).

\end{document}